\documentclass[12pt]{article}
\usepackage{amsfonts,amssymb,xspace}
\makeatletter
\def\address{\m@th\@ifnextchar[\@address{\@address[]}}

\def\@address[#1]#2{
\expandafter\def\expandafter\@addressname\expandafter
{\@addressname{
  \adr{#1}\ \parbox[t]{5in}{
     \ignorespaces #2}\par }}}
\def\@addressname{}
\def\adr#1{{\normalsize\unskip$^{#1}$}}
\def\@maketitle{%
\def\and{{\rm and}}
  \newpage
  \null
  \vspace*{-7ex}
  \flushright{hep-th/9612023}\\[4ex]
  {\centering
  \let \footnote \thanks
    {\Large\bf   \@title \par}%
    \vskip 1.5em%
      \lineskip .5em%
    {\bf\normalsize   \@author\par}
      \vspace{1em}
    {\small \@addressname}

  }%
  \par
  \vskip 1.5em}
\def\section{\@startsection {section}{1}{\z@}{-3.5ex plus-1ex minus
    -.2ex}{1.5ex plus.2ex}{\reset@font\large\bf}}
\def\subsection{\@startsection{subsection}{2}{\z@}{-3.25ex plus-1ex
    minus-.2ex}{1.5ex plus.2ex}{\reset@font\normalsize\bf}}
\def\subsubsection{\@startsection
     {paragraph}{4}{\z@}{3.25ex plus1ex minus.2ex}{-1em}{\reset@font
     \normalsize\bf}}
\makeatother
\title{A Nonlocal Transcendental Realization of\\
            the Sugawara Operators
            at Arbitrary Level\footnote{Talk presented at \emph{XXI
            International Colloquium on Group Theoretical Methods in
            Physics,\qquad \hspace*{5mm} 15 - 20 July 1996,
            Goslar, Germany}}}
	\author{Reinhold W. Gebert\footnote{Supported by Deutsche
	Forschungsgemeinschaft}}
	\address{Institute for Advanced Study,
                 School of Natural Sciences, \\
                 Olden Lane, Princeton, NJ 08540, U.S.A.}
\makeatletter
\def\sideset#1#2#3{%
  \setbox\z@\hbox{$\displaystyle{\vphantom{#3}}#1{#3}\m@th$}%
  \setbox\tw@\hbox{$\displaystyle{#3}#2\m@th$}%
  \hskip\wd\z@\hskip-\wd\tw@\mathop{\hskip\wd\tw@\hskip-\wd\z@
  {\vphantom{#3}}#1\!{#3}#2}}
\makeatother
\newcommand{\Cn}{\mathbb{C}}
\newcommand{\Rn}{\mathbb{R}}

\newcommand{\Zn}{\mathbb{Z}}

\newcommand{\lb}[1]{\label{#1}}
\newcommand{\Eq}[1]{(\ref{#1})}

\newcommand{\beq}{\begin{equation}}
\newcommand{\eeq}{\end{equation}}

\newcommand{\non}{\nonumber \\}

\newcommand{\ga}{\alpha}

\newcommand{\gd}{\delta}

\newcommand{\gl}{\lambda}
\newcommand{\gL}{\Lambda}

\newcommand{\gO}{\Omega}
\newcommand{\gp}{\psi}

\newcommand{\gr}{\rho}

\newcommand{\gx}{\xi}

\newcommand{\gz}{\zeta}
\newcommand{\cl}{\ensuremath{\ell}\xspace}

\newcommand{\cF}{\mathcal{F}}

\newcommand{\cP}[1][{}]{\mathcal{P}}
\newcommand{\cPL}{\ensuremath{\mathcal{P}(\vgL)}}
\newcommand{\cPl}{\ensuremath{\mathcal{P}^{(\svl)}}}
\newcommand{\cX}{\mathcal{X}}

\newcommand{\fg}{\ensuremath{\mathfrak{g}}\xspace}
\newcommand{\fh}{\mathfrak{h}}

\newcommand{\va}{\mathbf{a}}
\newcommand{\vga}{\mbox{\boldmath$\ga$}}

\newcommand{\vgd}{\mbox{\boldmath$\gd$}\xspace}

\newcommand{\vk}{\mathbf{k}}

\newcommand{\vgl}{\mbox{\boldmath$\gl$}\xspace}
\newcommand{\vgL}{\mbox{\boldmath$\gL$}\xspace}
\newcommand{\vo}{\mathbf{0}}
\newcommand{\vp}{\mathbf{p}}
\newcommand{\vP}{\mathbf{P}}
\newcommand{\vq}{\mathbf{q}}
\newcommand{\vQ}{\mathbf{Q}}
\newcommand{\vX}{\mathbf{X}}
\newcommand{\vr}{\mathbf{r}}

\newcommand{\vv}{\mathbf{v}}

\newcommand{\vro}{\mbox{\boldmath$\gr$}}
\newcommand{\vgx}{\mbox{\boldmath$\gx$}}

\newfont\sgb{cmmib8} 

\newcommand{\svd}{{\hbox{\sgb \symbol{"0E}}}}

\newcommand{\svl}{{\hbox{\sgb \symbol{"15}}}}
\newcommand{\svL}{{\hbox{\sgb \symbol{"03}}}}
\newcommand{\Oint}{\oint\limits}

\newcommand{\Res}[2][]{\Oint_{#1}\!\frac{d#2}{2\pi i}\,}
\renewcommand{\span}{\mathrm{span}}
\newcommand{\mult}{\mathrm{mult}}
\newcommand{\re}{\mathrm{e}}
\newcommand{\frc}[2]{{\textstyle \frac{#1}{#2}}}
\newcommand{\X}{\!\cdot\!}
\newcommand{\ket}[1]{\ensuremath{|#1\rangle}}
\newcommand{\ord}[1]{\mbox{\large\bf:} #1 \mbox{\large\bf:}}
\newcommand{\xord}[1]{{}_\times^\times #1 {}_\times^\times}
\newcommand{\hv}{\ensuremath{h^\vee}\xspace}
\newcommand{\bfg}{\ensuremath{\bar{\fg}}\xspace}

\newcommand{\bD}{\ensuremath{\bar{\Delta}}\xspace}
\newcommand{\vkl}{\ensuremath{\vk_\cl}\xspace}
\newcommand{\Ai}[1]{\ensuremath{A^i_{#1}}}

\newcommand{\Er}[1]{\ensuremath{E^{\vr}_{#1}}}
\newcommand{\Ema}[1]{\ensuremath{E^{-\vr}_{#1}}}
\newcommand{\XI}[1]{\ensuremath{\cX^i_{#1}}}
\newcommand{\vcX}{\ensuremath{\cX}}
\newcommand{\sL}[1]{\ensuremath{\mathcal{L}_{#1}}}
\newcommand{\tr}[2][\cl]{{\sideset{^{\scriptscriptstyle[#1]}}{_{#2}}{t}}}

\begin{document}
\maketitle
\begin{abstract}
A recently found new free field realization of the affine Sugawara
operators at arbitrary level is reviewed, which involves exponentials
of the well-known DDF operators in string theory.
\end{abstract}
\section{Results}
The Sugawara operators play an important role in the representation
theory of affine Lie algebras (see e.g.\ \cite{Kac90}). Using a string
vertex operator construction of the affine algebra at arbitrary level
it is possible to find a new free field realization of the Sugawara
operators at arbitrary level in terms of physical string ``DDF
oscillators''. This article reviews these results which were obtained
in collaboration with K.~Koepsell and H.~Nicolai. A complete set of
references can be found in \cite{GeKoNi96}.

Let \bfg be a finite-dimensional simple Lie algebra of type $ADE$ and
rank $d-2$ ($d\ge3$). Consider the associated nontwisted affine Lie
algebra \fg of rank $d-1$ with Cartan--Weyl basis $H^i_n$, $\Er{n}$ ($1\le
i\le d-2$, $\vr\in\bD$, $n\in\Zn$). The Sugawara operators
\beq  \sL{m}
   :=\frac1{2(\cl+\hv)}\sum_{n\in\Zn}
     \bigg(\sum_{i=1}^{d-2}\xord{H^i_nH^i_{m-n}}
           +\sum_{\vr\in\bD}\xord{\Er{n}\Ema{m-n}}\bigg) \lb{Sug0} \eeq
then form a Virasoro algebra with central charge
$c_\cl:=\frac{\cl\dim\bfg}{\cl+\hv}$, where $\cl$ and $\hv$ denote the
level and the dual Coxeter number, respectively. Our main result is
that in a certain string model the operators can be rewritten in terms
of free oscillators \Ai{m}:
\begin{eqnarray}
  \sL{m}
    &=&\frac1{2\cl}\sum_{n\in\Zn}
       \sum_{i=1}^{d-2}\xord{\Ai{\cl n}\Ai{\cl(m-n)}}
       +\frac{\hv}{2\cl(\cl+\hv)}\sum_{n\neq 0(\cl)}
              \sum_{i=1}^{d-2}\xord{\Ai{n}\Ai{\cl m-n}} \non
    & &{}-\frac1{2\cl(\cl+\hv)}\sum_{\vr\in\bD}
              \sum_{p=1}^{\cl-1}\frac1{|\gz^p-1|^2}
              \Oint\!\frac{dz}{2\pi i}\,z^{\cl m -1}
              \xord{\re^{i\vr\cdot \left[\vcX(z_p)-\vcX(z)\right]}} \non
    & &{}+\frac{(\cl^2-1)(d-2)\hv}{24\cl(\cl+\hv)}\gd_{m,0},
   \lb{Sug}
\end{eqnarray}
where $\gz:=\re^{2\pi i/\cl}$, $z_p:=\gz^pz$,
and
\beq  \XI{}(z):= Q^i- i\Ai{0}\ln z+
              i\sum_{m\ne0}\frac1m\Ai{m}z^{-m}. \lb{Fub} \eeq
(The zero modes $Q^i$ apparently drop out in \Eq{Sug} but will be
needed later.) It will turn out that the ``DDF oscillators'' \Ai{m} are
constructed from exponentials of the ordinary string oscillators
$\ga^\nu_n$. Therefore the $\sL{m}$'s are in fact ``doubly
transcendental'' functions of the string oscillator modes. It is easy
to see that for the special case $\cl=1$ the Sugawara operators take
the well-known form
\beq  \sL{m}=\frac12\sum_{n\in\Zn}\sum_{i=1}^{d-2}\xord{H^i_nH^i_{m-n}}, \eeq
which is nothing but the equivalence of the Virasoro and the Sugawara
construction at level 1 proved by I.~Frenkel. On the other hand, if we
consider the action of the zero mode operator on an arbitrary
level-\cl affine highest weight vector $\ket\vgL$, then we obtain
(after invoking some identity for sums over roots of unity) the result
\beq  \sL0\ket\vgL=\frac{(\bar{\vgL}+2\bar{\vro})\X\bar{\vgL}}%
                       {2(\cl+\hv)}\ket\vgL, \eeq
which had previously been derived by exploiting the properties of the
affine Casimir operator (see e.g.\ \cite{Kac90}), whereas here it can
be simply read off from the general formula as a special
case. Finally, formula \Eq{Sug} exhibits some nonlocal structure due
to a new feature in the operator product expansion. In conformal field
theory the singular part of the operator product expansion usually
involves negative powers of $z-w$ leading to poles at $z=w$. In our
case, however, we will demonstrate the appearance of negative powers
of $z^\cl-w^\cl$ which produces poles at $z=w_p:=\re^{2\pi i/\cl}w$
and is the origin of the nonlocal expressions in \Eq{Sug}.

\section{Compactified Bosonic String and DDF Construction}
The string model in which we realize the affine Lie algebra and the
Sugawara operators is ``finite in all directions'', i.e., we consider
a (chiral half of a) closed bosonic string moving on a $d$-dim
Minkowskian torus as spacetime such that the momentum lattice is given
by the affine weight lattice $Q^*$. The usual string oscillators
$\ga^\mu_m$ ($1\le\mu\le d$, $m\in\Zn$) form a $d$-fold Heisenberg
algebra $\mathbf{h}$,
$[\ga^\mu_m,\ga^\nu_n]=m\eta^{\mu\nu}\gd_{m+n,0}$,
and one introduces groundstates $\ket\vgl=\re^{i\svl\cdot \vq}\ket\vo$ for
$\vgl\in Q^*$, which are by definition highest weight states for
$\mathbf{h}$, i.e., $\ga^\mu_0\ket\vgl=\gl^\mu\ket\vgl$ and
$\ga^\mu_m\ket\vgl=0\quad\forall m>0$. ($q^\nu$ is position
operator:$[q^\mu,p^\nu]=i\eta^{\mu\nu}$, $p^\mu\equiv\ga^\mu_0$) Then
the Fock space $\cF$ is the direct sum of irreducible $\mathbf{h}$
modules:
$\cF=\span\{\ga^{\mu_1}_{-m_1}\cdots\ga^{\mu_M}_{-m_M}\ket\vgl\,|\,
              1\le\mu_i\le d, m_i>0, \vgl\in Q^*\}$.
To complete the quantization procedure one has to implement the
physical state conditions. This amounts to restricing $\cF$ to the
subspace $\cP$ of physical states, which are by definition conformal
primary states of weight 1, viz.
\beq  \cP:=\bigoplus_{\svl\in Q^*} 
        \cPl,\qquad
   \cPl:=\{\gp\in\cF\,|\,
   L_n\gp=\gd_{n0}\gp\ \forall n\ge0,\
   p^\mu\gp=\gl^\mu\gp\}, \eeq
with respect to the Virasoro constraints (with $c=d$)
$L_n:=\frac12\sum_{m\in\Zn}\ord{\vga_m\X\vga_{n-m}}$. One easily
works out the simplest examples of physical string states to find
tachyons $\ket\va$, satisfying $\va^2=2$, and photons
$\vgx\X\vga_{-1}\ket\vk$, satisfying $\vgx\X\vk=\vk^2=0$
($\vgx\in\Rn^{d-1,1}$). This direct method quickly becomes rather
cumbersome and one might ask whether there is an elegant way of
describing physical states which also yields structural insights into
$\cP$. This is achieved by the so-called DDF construction.

Let us consider a fixed momentum vector $\vgl\in Q^*$ satisfying
$\vgl^2\le2$ (otherwise it could not give rise to physical states). In
order to find a complete basis for $\cPl$, one starts from a
so-called DDF decomposition of $\vgl$,
\beq  \vgl=\va-n\vk, \qquad n=1-\frc12\vgl^2\ , \eeq
for some tachyon $\ket\va$ and lightlike vector $\vk$ satisfying
$\va\X\vk=1$. Such a decomposition is always possible although neither
$\va$ nor $\vk$ will in general lie on the affine weight lattice.
Next we choose $d-2$ orthonormal polarization vectors $\vgx^i\in\Rn^{d-1,1}$
such that $\vgx^i\X\va=\vgx^i\X\vk=0$. The transversal DDF operators
are now defined as
\beq  \Ai{m}(\va,\vk)
    := \Res{z}\vgx^i\X\vP(z)\,
                \re^{im\vk\cdot \vX(z)}, \eeq
with the Fubini--Veneziano coordinate and momentum fields respectively
given by
\beq  X^\mu(z)
   := q^\mu-ip^\mu\ln z+i\sum_{m\ne0}\frac1m\ga^\mu_mz^{-m},\qquad
   P^\mu(z)
   := \sum_{m\in\Zn}\ga^\mu_mz^{-m-1}. \eeq
It is straightforward to show that the $\Ai{m}$'s realize a
$(d-2)$-fold ``transversal'' Heisenberg algebra,
$[A^i_m,A^j_n]=m\gd^{ij}\gd_{m+n,0}$.  One also introduces
longitudinal DDF operators $A^-_m(\va,\vk)$, which are much more
complicated expressions (and whose explicit form is not needed
here). They form a ``longitudinal'' Virasoro algebra ($c=26-d$) and
satisfy $[A^i_n,A^-_m]=0$. It turns out that the DDF operators
constitute a spectrum-generating algebra for $\cPl$ in the following
sense. First, one can show that $[A^i_n,L_m]=[A^-_n,L_m]=0\ \forall
m$, which implies that they map physical states into physical
states. Furthermore, the tachyonic state $\ket\va$ is annihilated by
the DDF operators with nonnegative mode index. Finally, the DDF
operators provide a basis for $\cPl$, viz.
\beq  \cPl=\span\{ A^{i_1}_{-n_1}\cdots A^{i_N}_{-n_N}
   A^-_{-m_1}\cdots A^-_{-m_M}\ket\va\,|\,
   n_1+\ldots+m_M=1-\frc12\vgl^2\}. \eeq

\section{Realization of Affine Lie Algebra}
We will now employ a vertex operator construction for the affine Lie
algebra at arbitrary level to find an explicit realization of the
Sugawara operators in terms of the string oscillators. Let
$L(\vgL)$ denote an irreducible level-\cl affine highest weight module
with vacuum vector $v_{\svL}$, dominant integral weight $\vgL\in Q^*$,
and weight system $\gO(\vgL)$. Without loss of generality we may
assume that $\vgL^2=2$ (due to $L(\vgL)\cong L(\vgL+z\vgd)$ for
$z\in\Cn$ and the affine null root $\vgd$). We define a Cartan--Weyl
basis for \fg by
\beq  H^i_m
    := \Res{z}\vgx^i\X\vP(z)\,\re^{im\svd\cdot \vX(z)}, \qquad
   \Er{m}
    := \Res{z}\ord{\re^{i(\vr+m\svd)\cdot \vX(z)}}c_\vr, \eeq
where $c_\vr$ is some cocycle factor. The central element and the
exterior derivative are given by $K:=\vgd\X\vp$ and $d:=\vgL_0\X\vp$,
respectively.  This yields a level-\cl vertex operator realization of
\fg on the space of physical states with $v_{\svL}\equiv\ket\vgL$,
viz.
\beq   L(\vgL)\hookrightarrow\cPL:=\bigoplus_{\svl\in\gO(\svL)}\cPl,\qquad
    L(\vgL)_{\svl}\hookrightarrow\cPl. \eeq
Note that the operators $\Ai{\cl m}\equiv H^i_m$ are not only part of
the transversal Heisenberg algebra but also make up the homogeneous
Heisenberg subalgebra of \fg. Only at level 1 these two algebras
coincide. It is intriguing to see how in the above construction both
the vacuum vector conditions $e_I\ket{\vgL}=0$ and the null vector
conditions $f_I^{1+\vr_I\cdot \svL}\ket{\vgL}=0$ for $0\le I\le d-2$ (in
terms of the affine Chevalley generators $e_I,f_I$) immediately follow
from the physical state condition. Below, we will see that only
transversal physical states can occur in the affine highest weight
module $L(\vgL)$. Hence we effectively deal with the embedding
$L(\vgL)_{\svl}\hookrightarrow \cPl_{\mathrm{transv.}}$ and have the
following universal estimate for affine weight multiplicities at
arbitrary level:
\beq  \mult_{\svL}(\vgl)\equiv\dim L(\vgL)_{\svl}
    \le \dim \cPl_{\mathrm{transv.}}=p_{d-2}(1-\frc12\vgl^2), \eeq
where $p_{d-2}(n)$ counts the partition of $n$ into ``parts'' of $d-2$
``colours''.

Finally, we would like to sketch how the new formula \Eq{Sug} is
obtained. If we insert the above expressions for the step operators
$\Er{n}$ into \Eq{Sug0}, we encounter in
$\sum_{n\in\Zn}\xord{\Er{n}\Ema{m-n}}$ an operator-valued infinite
series
\beq  Y(z,w) := \sum_{n\ge0}\re^{in\svd\cdot [\vX(z)-\vX(w)]}. \eeq
It is not difficult to see that application to the groundstate
$\ket\va$ yields
\beq  Y(z,w)\ket\va
   = \sum_{n\ge0}\left[\left(\frac{z}{w}\right)^{\cl n}
                          +\mbox{osc.\ terms}\right]\ket\va
   = \frac{z^\cl}{z^\cl-w^\cl}\ket\va + \mbox{excited
      states}, \eeq
which explicates the origin of the nonlocalities. Another important
point to notice is that the Sugawara operators are physical by
construction. Consequently, it must be possible to rewrite them in a
manifestly physical form, i.e., in terms of the DDF oscillators. At
first sight this seems to be a hopeless task but there is a little
trick. Motivated by the observation that the leading oscillator
contribution in a DDF operator $\Ai{m}$ is given by $\vgx^i\X\vga_m$,
one simply replaces every oscillator term $\vr\X\vga_m$ occurring in
the vertex operator realization of the Sugawara operators by a ``DDF
oscillator'' $(\vr\X\vgx^i)\Ai{m}$. The resulting formula \Eq{Sug}
then indeed gives an alternative expression for the Sugawara operators
in terms of the transversal DDF operators \cite{GeKoNi96}.

This method also works for the step operators themselves
\cite{GebNic97}. One ends up with
\beq  \Er{m}\Big|_{\cP(\svL)} = \Res{z} z^{\cl m}
  \xord{\re^{i\vr\cdot \vcX(z)}}c_\vr, \eeq
with the new transversal coordinate field $\vcX(z)$ given in
\Eq{Fub}. There is one subtlety here concerning the zero modes. Since
the position operators $q^\mu$ are unphysical we have to replace them
by some physical analogue. What should that be? The correct answer
involves the (physical!) Lorentz generators $M^{\mu\nu}$, viz.
\beq  Q^i :=(\vgx^i)_\mu(\vkl)_\nu M^{\mu\nu}. \eeq
One shows that
$\re^{i\vr\cdot \vQ}\vv\X\vga_m\re^{-i\vr\cdot\vQ} =\tr{\vr}(\vv)\X\vga_m$,
where $\tr{\vr}(\vv) := \vv + (\vv\X\vkl)\vr -
\left[(\vv\X\vkl)\frc12\vr^2 + \vr\X\vv\right]\vkl$ for $\vv\in\fh^*$,
$\vkl:=\frc1\cl\vgd$, $\vr\in\bar{Q}$. Hence the resolution of the
zero mode problem provides us with a realization of the well-known
affine Weyl translations $\tr{\vr}$ in terms of Lorentz boosts.

How will the longitudinal DDF operators fit into this framework?
Clearly, the full space of physical states is an infinite direct sum
of irreducible \fg-modules. Then the longitudinal DDF operators will
map between different levels or, more generally, between different
\fg-modules and will thus play the role of intertwining operators.

\end{document}